**Unexpected composition dependence of the first sharp diffraction peak in an alcohol-aldehyde liquid mixture: *n*-pentanol and pentanal**

*Ildikó Pethes\*, László Temleitner, Matija Tomšič, Andrej Jamnik, László Pusztai*

I. Pethes, L. Temleitner, L. Pusztai
Wigner Research Centre for Physics, Hungarian Academy of Sciences,
Budapest, Konkoly Thege út 29-33., H-1121, Hungary

L. Pusztai
International Research Organisation for Advanced Science and Technology (IROAST), Kumamoto University, 2-39-1 Kurokami, Chuo-ku, Kumamoto 860-8555, Japan

M. Tomšič, A. Jamnik
Faculty of Chemistry and Chemical Technology (FKKT), University of Ljubljana,
Večna pot 113, 1000 Ljubljana, Slovenia

Corresponding author: e-mail: pethes.ildiko@wigner.mta.hu

**Abstract**

The total scattering structure factors of pure liquid *n*-pentanol, pentanal, and 5 of their mixtures, as determined by high energy synchrotron X-ray diffraction experiments, are presented. For the interpretation of measured data, molecular dynamics computer simulations are performed, utilizing 'all-atom' type force fields. The diffraction signals in general resemble each other over most of the monitored $Q$ range above 1 Å$^{-1}$, but the absolute values of the intensities of the small-angle scattering maximum ('pre-peak', 'first sharp diffraction peak'), around 0.6 Å$^{-1}$, change in an unexpected fashion, non-linearly with the composition. MD simulations are not able to reproduce this low-$Q$ behavior; on the other hand, they do reproduce the experimental diffraction data above 1 Å$^{-1}$ rather accurately. Partial radial distribution functions are calculated based on the atomic coordinates in the simulated configurations. Inspection of the various O-O and O-H partial radial distribution functions clearly shows that both the alcoholic and the aldehydic oxygens form hydrogen bonds with the hydrogen atoms of the alcoholic OH-group.

1. Introduction

Reactions between alcohols and aldehydes/ketones are well-known in organic chemistry (see, e.g., Ref. [1]): under appropriate conditions (e.g., acidic environment, catalyst, necessity of removing water during the reaction, etc…), they lead to the formation of acetals. Acetals can be found in various applications – for example, as solvents in the chemical industry, protective agents for the aldehyde group during multi-step syntheses. Interestingly, most glycosidic bonds in carbohydrates are acetalic linkages.[1]

When alcohols and aldehydes are simply mixed (under no special conditions), a fully reversible reaction between the aldehyde -CHO and the alcoholic –OH groups can lead to the formation of hemiacetals, R-C[HO][OR'], where R and R' represent alkyl groups. Hemiacetals cannot be separated from such liquid mixtures (containing an alcohol and an aldehyde only) – their presence may only be noticed indirectly (e.g., by spectroscopic methods, such as NMR).[1]

In this study we wished to establish whether any trace of hemiacetal formation can possibly be detected via diffraction methods. For this purpose, medium sized aliphatic alkyl chains have been selected; the final choices were *n*-pentanol as the alcohol part, and pentanal (pentane-aldehyde) as the aldehyde. The components of our liquid mixtures can be viewed in **Figure 1**.

The two pure components and 5 mixtures over the entire concentration range have been taken to an X-ray diffractometer at a synchrotron source for determining the total scattering structure factor over a wide range of the scattering variable. For interpreting diffraction data, molecular dynamics (MD) computer simulations have been conducted. Details of these investigations, as well as results and their discussions, are provided in the following sections.

**2. Experimental**

Five pentanal/*n*-pentanol mixtures were investigated, with pentanal concentrations of 90, 75, 50, 25 and 10 mol%, together with pure pentanal and *n*-pentanol.

X-ray diffraction measurements have been carried out at the BL04B2[2] high energy X-ray diffraction beamline of the Japan Synchrotron Radiation Research Institute (SPring-8, Hyogo, Japan), using a wavelength of 0.2023 Å (beam energy: 61.3 keV). This way, we were able to record the diffraction patterns of the samples in transmission mode, in the horizontal scattering plane, using a single HPGe detector between scattering variable, $Q$, values of 0.16 and 16 Å$^{-1}$. Samples were contained in 2 mm diameter, thin-walled quartz capillaries (GLAS Müller, Germany). The capillaries were mounted in the automatic sample changer of the BL04B2 instrument. In order to prevent the evaporation of samples during the measurement, the capillaries were sealed with a cap created from folded Parafilm(R). Diffraction patterns were recorded in three overlapping frames that differed by the width of incoming beam.

Diffracted intensities were normalized by the incoming beam monitor counts, corrected for absorption, polarization and contributions from the empty capillary. Finally, the patterns over the entire Q-range were obtained by normalizing and merging each frame in electron units, then removing inelastic (Compton) scattering contributions following a standard procedure.[3]

## 3. Molecular dynamics simulations

Classical molecular dynamics (MD) simulations were carried out by using the GROMACS software package (version 2016.3).[4] The calculations were performed at constant volume and temperature (NVT ensemble), at T = 293.15 K. Cubic simulation boxes were applied, with periodic boundary conditions. The boxes contained 2000 molecules in total for all studied mixtures; the numbers of pentanal and *n*-pentanol molecules are given in **Table 1**. Densities of the mixtures were estimated using the densities of the pure pentanal and *n*-pentanol. Simulation box sizes were calculated according to the densities; these are also shown in Table 1.

For the description of interatomic interactions the OPLS-AA force field[5] was applied. In this model all atoms are treated explicitly, as shown in Figure 1. Non-bonded interactions are described by the 12-6 Lennard-Jones (LJ) interaction and the Coulomb potential. The LJ parameters ($\varepsilon_{ii}$ and $\sigma_{ii}$) and the partial charges applied ($q_{ii}$) to the different atoms are collected in **Table 2**. The $\varepsilon_{ij}$ and $\sigma_{ij}$ parameters between unlike atoms are calculated as the geometric average of the homoatomic parameters (geometric combination rule). The intramolecular non-bonded interactions between pairs of atoms separated by three bonds (1-4 interactions) are reduced by a factor of 2; between first and second neighbors (atoms separated by one or two bonds) these interactions are neglected. Intramolecular bonded forces are calculated in the form of bond stretching, angle bending and dihedral angle torsion.

Bond lengths within pentanal and *n*-pentanol molecules were fixed (using the LINCS algorithm[6]), while bond angles and torsional angles were flexible. Bond lengths, equilibrium angles and force constants are given in **Table 3, 4 and 5**. The smoothed particle-mesh Ewald (SPME) method[8] was used for treating Coulomb interactions, using a 20 Å cutoff in direct space. The LJ interactions were cut-off at 20 Å.

Initially, molecules were placed in the simulation box randomly. At first an energy minimization was performed using the steepest-descent method. The equations of motions were integrated via the leapfrog algorithm; the time step was 2 fs. The temperature was maintained by a Nose-Hoover thermostat[9] with $\tau_T$=2.0 ps. After 2 ns of equilibration, 50 particle configurations, 40 ps apart, were collected.

Partial radial distribution functions (PRDF, $g_{ij}(r)$) were calculated from the collected configurations using the "gmx_rdf" programme of the GROMACS software. The model structure factor can be obtained from the PRDFs by the following equations (Equation (1) and (2)):

$$S_{ij}(Q) - 1 = \frac{4\pi\rho_0}{Q} \int_0^\infty r\big(g_{ij}(r) - 1\big)\sin(Qr)\mathrm{d}r \qquad (1)$$

$$F_{\mathrm{mod}}(Q) = \sum_{i \leq j} w_{ij}(Q) S_{ij}(Q), \qquad (2)$$

where $Q$ is the amplitude of the scattering vector, $\rho_0$ is the average number density. The $w_{ij}(Q)$ scattering weighting factors can be obtained by Equation (3):

$$w_{ij}(Q) = (2 - \delta_{ij}) \frac{c_i c_j f_i(Q) f_j(Q)}{\sum_{ij} c_i c_j f_i(Q) f_j(Q)}. \qquad (3)$$

Here $\delta_{ij}$ is the Kronecker delta, $c_i$ denotes the atomic concentration and $f_i(Q)$ is the atomic form factor of atom type $i$; this latter quantity may be calculated by a parametrized form.[10]

### 4. Results and discussion

Corrected and normalized experimental total scattering structure factors from X-ray diffraction are shown in **Figure 2**, together with the X-ray weighted counterparts calculated from molecular dynamics simulation results. The overall agreement, as seen in direct comparisons (Figure 2), appears acceptable, and MD calculations are apparently able to reproduce the trends as composition changes – above the value of the scattering vector of 1 Å$^{-1}$. On this basis, it may be stated that MD calculations are able to reproduce the structural trends in this mixture as its composition changes at the level of short correlation lengths reasonably well. For example, the varying absolute value of the intensity of the main maximum is in the same order for the calculated $F(Q)$-s as they are for the experimental curves, see **Figure 3**. Similarly, simulated intensities around the minimum after the main peak, as well as around maxima at higher $Q$-values behave correctly. The only noticeable discrepancy between the experimental and simulation data occurs above the $Q$ value of 1 Å$^{-1}$, where a slight shift in terms of the position of the main maximum is observed. This shift does not show up for pure *n*-pentanol but as soon as pentanal is added, the discrepancy comes into view. Based on the fresh MD simulation study of simple aldehydes,[11] where the same kind of shift was observed for pure aldehydes from propanal to nonanal, it may be claimed that this is due to non-perfect pentanal potential parameters.

However, the really remarkable and unexpected phenomenon is to be found at $Q$ values below 1 Å$^{-1}$, see Figure 2 and 3. Pure *n*-pentanol shows a sizeable pre-peak, i.e. first maximum in the low-$Q$ regime, with an intensity of about 1/5 of that of the main maximum, whereas pure pentanal shows only a shallow scattering shoulder in this small-angle regime. MD simulations predict that, according to commonsense expectations, the intensity of the pre-peak would decrease monotonically from *n*-pentanol to pentanal. Experimental results, however, refute this expectation: the intensity of the small-angle maximum monotonically grows from *n*-pentanol to that of the mixture with 25 % pentanal, and for the mixture with 50 % pentanal, the pre-peak is still higher than it is for pure *n*-pentanol. At the same time, the intensity of the main maximum changes only moderately: it's decreasing monotonically from *n*-pentanol to pure pentanal. It can be stated that $F(Q)$-s hardly change at all, as a function of composition, above the scattering variable value of 2.5 Å$^{-1}$.

The OPLS-AA type interatomic potentials do not seem to be able to reproduce this most interesting feature. One can therefore only speculate at this stage that what diffraction data can detect at low $Q$ values is actually the signature of the reversible chemical reaction mentioned in the Introduction. The fact is that an influence of such reaction on the structure could not be observed in the simulated data as such reaction is not modeled, but it would certainly be expected to influence the experimental data on the real mixtures. This presumption may perhaps be checked by running Reverse Monte Carlo (RMC) type[12] calculations, where it is guaranteed (for reasonably good data at least) that the structural model provided by RMC would reproduce the measured total scattering structure factors. By analyzing final RMC configurations, it may be possible to find hints concerning what causes the unexpected behavior of the pre-peak in these mixtures.

Since agreement between experiment and MD simulations is excellent above $Q$ approx. 1 Å$^{-1}$, in what follows, we wish to analyze some of the partial radial distribution functions that have been calculated directly from the simulated atomic coordinates. It is generally believed that the low-$Q$ behavior of the total scattering structure factors reflects characteristics over larger length-scales, of the order of nm (frequently referred to as 'intermediate range ordering', IRO).[13] Therefore, interpreting structural results on the atomic scale can be justified. Our main interest here is hydrogen bonding (H-bonding) that certainly is present between molecules of *n*-pentanol; the question is whether H-bonds can form between *n*-pentanol and pentanal molecules (the alcoholic –OH would be the proton donor in such a formation).

H-bond-related partial radial distribution functions for one selected mixture, the one that contains 50-50 % of alcohol and aldehyde, are shown in **Figure 4**. For this composition, the intensity of the small-angle scattering maximum ('pre-peak') of the total $F(Q)$ is still higher than that of the similar peak of pure *n*-pentanol. Positions of the first maxima for the PRDF-s displayed are collected in **Table 6**.

There are 3 different O-O atom pairs possible in the system: alcohol-O – alcohol-O (O-O in short), alcohol-O – aldehyde-O (O-O') and aldehyde-O – aldehyde-O (O'-O'). Of the many possible O-H PRDF-s, four of them are relevant from the point of view of possible H-bonding: alcohol-O – alcohol-H(OH) (O-H$_O$ in short), alcohol-O – aldehyde-H'(COH) (O-H'), aldehyde-O – alcohol-H(OH) (O'-H$_O$) and aldehyde-O – aldehyde-H'(COH) (O'-H'). As it may be evident from Figure 4, there is a very sharp distinction between the various O-O PRDF-s: if one of the participating oxygen atom is from the alcohol then the characteristic H-bonding O-O distance (of about 2.8 to 3.0 Å) appears.

Considering the relevant O-H PRDF-s, the pattern becomes a little even clearer: both alcohol and aldehyde oxygen atoms can bond with the alcohol-H (OH), at around 1.8 Å, whereas the aldehyde-H' (COH) cannot get closer to any of the oxygen atoms than about 2.5 Å (peak position: 2.8 Å). It follows that the presence of alcohol-aldehyde hydrogen bonds is strongly supported by the present study.

The composition dependence of some of the PRDF-s related to hydrogen bonding can be followed in **Figure 5**. The curves follow the patterns shown in Figure 4, part (a). No changes in terms of any of the maximum positions can be observed over the entire concentration range, which is quite the opposite of what happens to the pre-peak intensities (cf. Figure 3). That is, the unexpected behavior of the total scattering structure factors at low $Q$ cannot be explained simply by changes in terms of the hydrogen bonding characteristics.

## 5. Summary and conclusions

The total scattering structure factors of pure *n*-pentanol, pentanal, and their mixtures at five different concentrations of 10, 25, 50, 75 and 90 mol % of pentanal, have been determined by high energy synchrotron X-ray diffraction experiments over the scattering variable range between 0.3 and 16 Å$^{-1}$.

For the interpretation of measured data, molecular dynamics computer simulations have been performed on systems of 2000 molecules, using 'all-atom' type force fields – the hydrogen atoms were also explicitly considered.

The diffracted signals resemble each other over most of the $Q$ range, above 1 Å$^{-1}$. However, the intensity of the experimentally obtained pre-peak (small angle maximum), positioned at around 0.6 Å$^{-1}$, changes in an unexpected fashion, starting from *n*-pentanol: the intensity of this feature is increasing first, and only above a pentanal content of 75 mol % it decreases back to below the initial peak intensity. MD simulations were not able to reproduce this low-$Q$ behavior, which means that these simulation results cannot convincingly account for changes in terms of the supramolecular structure in the studied samples.

On the other hand, MD calculations successfully reproduce measured diffraction data above 1 Å$^{-1}$ rather accurately, which provides confidence for looking at the atomic scale structure. Partial radial distribution functions have been calculated from atomic coordinates in the simulated configurations.

Inspection of the various O-O and O-H PRDF-s made it clear that both the alcoholic and the aldehydic oxygens can form hydrogen bonds with the hydrogen atoms of the alcoholic OH-group, whereas the H atom of the aldehyde group cannot participate in H-bonding.

In order to clarify the unexpected low-$Q$ behavior of the measured structure factors, it might be beneficial to conduct similar investigations also for additional alcohol-aldehyde pairs. The interpretation of experimental data may be completed by conducting also Reverse Monte Carlo type computer modeling.

**Acknowledgment**

The authors are grateful to the National Research, Development and Innovation Office (NKFIH) of Hungary for financial support through Grants No. SNN 116198, TÉT_16-1-2016-0056, and to the Slovenian Research Agency for the financial support through the research core funding No. P1-0201 and the project No. N1-0042 (Structure and thermodynamics of hydrogen-bonded liquids: from pure water to alcohol-water mixtures). LP thanks Mr. Á. Pusztai (University of Cambridge) for useful discussions concerning organic chemistry aspects.

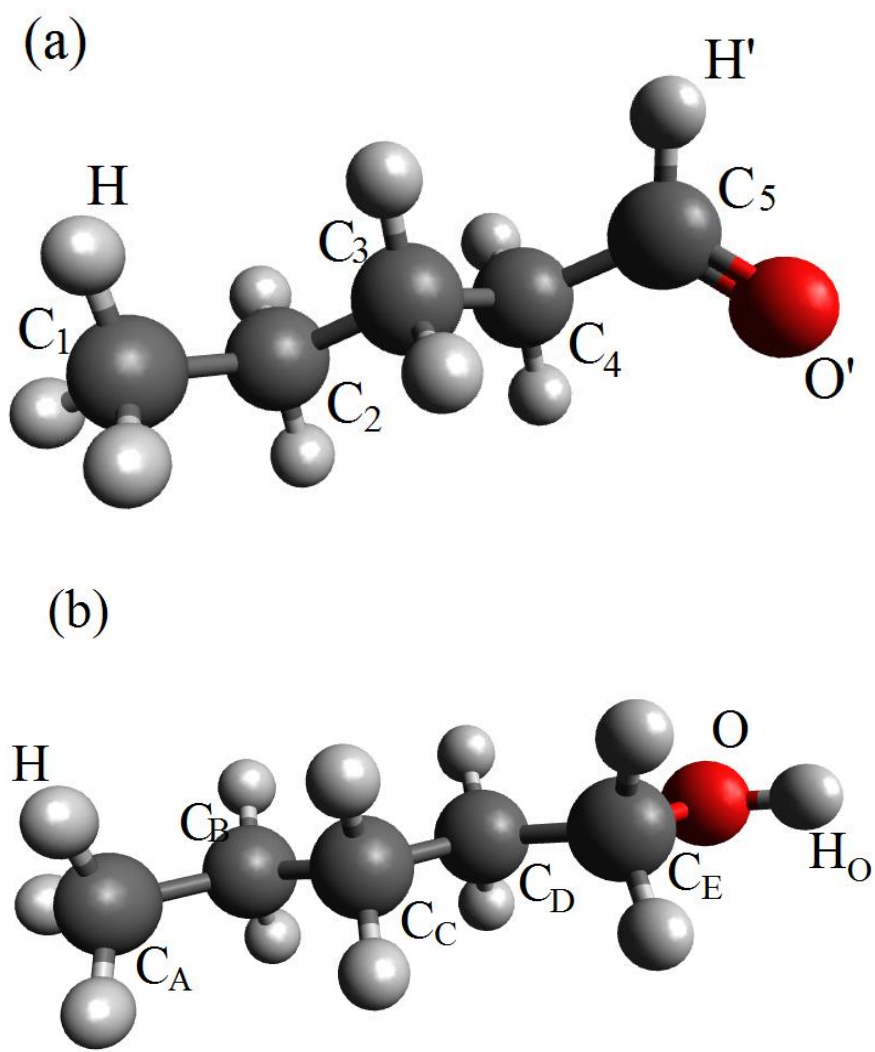

**Figure 1.** Schematic representation of the all-atom model of (a) pentanal and (b) *n*-pentanol molecules.

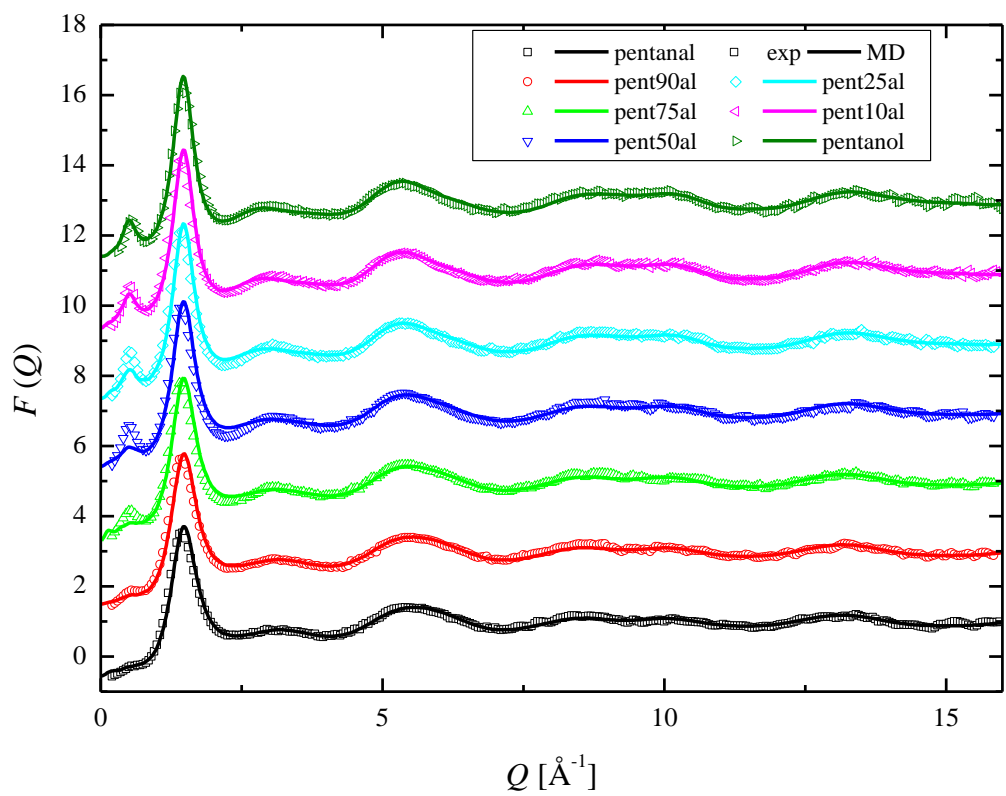

**Figure 2.** Experimental (symbols) and simulated (lines) X-ray diffraction structure factors of pentanal/*n*-pentanol mixtures. The curves are shifted upwards to improve clarity.

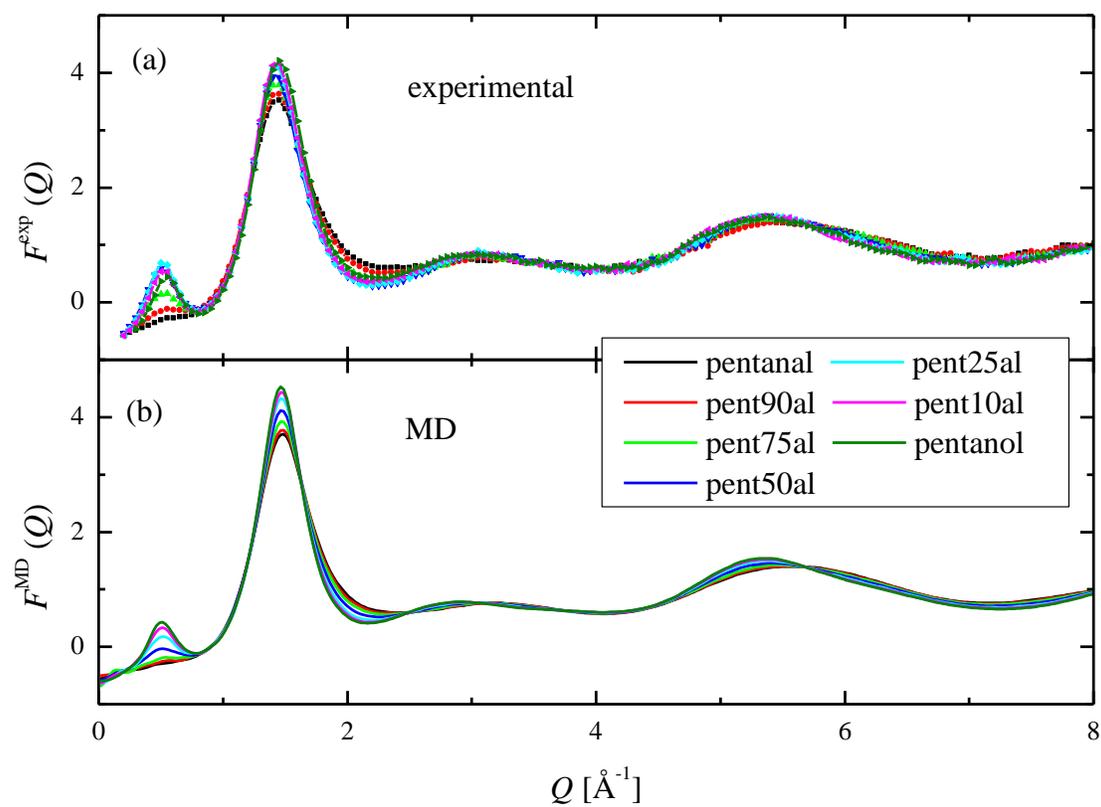

**Figure 3.** (a) Experimental and (b) simulated X-ray diffraction structure factors of the various pentanal/*n*-pentanol mixtures. The focus here is on the clearest presentation of the changes across the entire composition range.

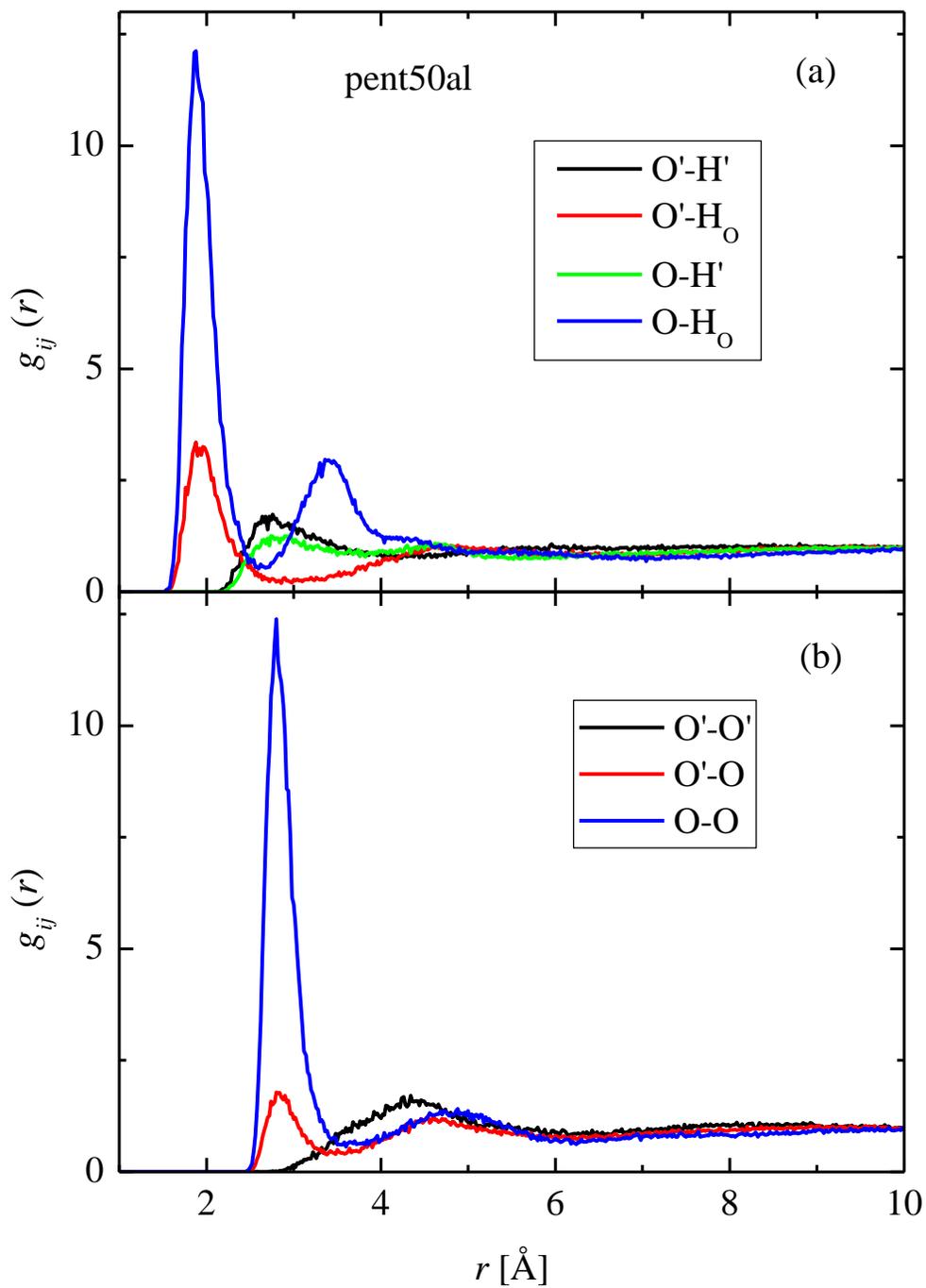

**Figure 4.** Intermolecular partial radial distribution functions of the 'pent50al' sample (50-50 mol% of both components) obtained from the corresponding MD simulation: (a) O'-H'(black), O'-$H_O$ (red), O-H' (light green) and O-$H_O$ (blue), and (b) O'-O' (black), O'-O (red) and O-O (blue).

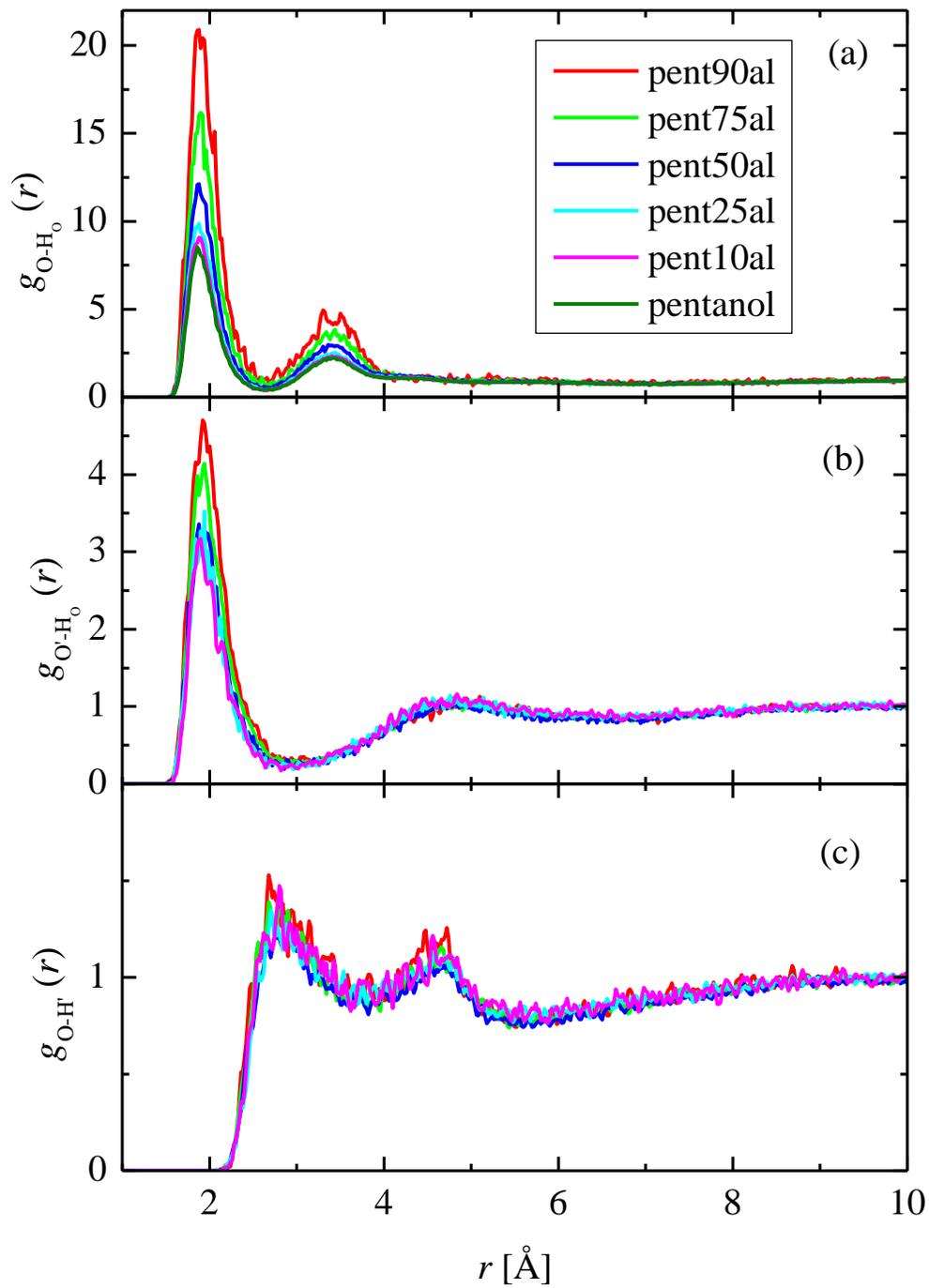

**Figure 5.** (a) Intermolecular O-H$_O$, (b) O'-H$_O$ and (c) O-H' partial radial distribution functions of pentanal/*n*-pentanol mixtures obtained from MD simulations.

**Table 1.** Idenfication of the investigated pentanal − *n*-pentanol mixtures, along with their densities, number densities, the numbers of pentanal and *n*-pentanol molecules in the simulation boxes, and box sizes. The total number of molecules was 2000 in all mixtures.

| Short name | $N_{pentanal}$ | $N_{n\text{-}pentanol}$ | Density [g cm$^{-3}$] | Number density [Å$^{-3}$] | Box length [nm] |
|---|---|---|---|---|---|
| pentanal | 2000 | 0 | 0.81 | 0.09061 | 7.06847 |
| pent90al | 1800 | 200 | 0.8101 | 0.09154 | 7.07368 |
| pent75al | 1500 | 500 | 0.81025 | 0.09293 | 7.08149 |
| pent50al | 1000 | 1000 | 0.8105 | 0.09522 | 7.09446 |
| pent25al | 500 | 1500 | 0.81075 | 0.09749 | 7.10738 |
| pent10al | 200 | 1800 | 0.8109 | 0.09883 | 7.11511 |
| pentanol | 0 | 2000 | 0.811 | 0.09973 | 7.12026 |

**Table 2.** Non-bonded parameters (partial charges and LJ parameters) used in MD simulations.[5] Atom type names are used in the bond, angle and dihedral types. Notation of atoms is shown in Figure 1.

| Atom | Type name | $q$ [e] | $\sigma_{ii}$ [nm] | $\varepsilon_{ii}$ [kJ mol$^{-1}$] |
|---|---|---|---|---|
| $C_2$, $C_3$, $C_4$, $C_B$, $C_C$, $C_D$ | C | -0.12 | 0.35 | 0.276144 |
| $C_1$, $C_A$ | C | -0.18 | 0.35 | 0.276144 |
| $C_E$ | C | 0.145 | 0.35 | 0.276144 |
| $C_5$ | C' | 0.45 | 0.375 | 0.43932 |
| O | O | -0.683 | 0.312 | 0.71128 |
| O' | O' | -0.45 | 0.296 | 0.87864 |
| H | H | 0.06 | 0.25 | 0.12552 |
| $H_O$ | $H_O$ | 0.418 | 0 | 0 |
| H' | H' | 0.0 | 0.242 | 0.06276 |

**Table 3.** Equilibrium bond lengths used in MD simulations. (Atom type names are shown in Table 2.)

| Bond type | Bond length [nm] |
|---|---|
| C-C | 0.1529 |
| C'-C | 0.1522 |
| C-H | 0.109 |
| C'-H' | 0.109 |
| C-O | 0.141 |
| C'-O' | 0.1229 |
| O-$H_O$ | 0.0945 |

**Table 4.** Angle bending parameters. Bond angle vibrations are represented by harmonic potentials, $\theta^0_{ijk}$ is the equilibrium angle and $k^a_{ijk}$ is the force constant. (For parameters that are not detailed in the original paper of Jorgensen et al.[5], values provided together with GROMACS package were used. Atom type names are shown in Table 2.)

| Angle type | $\theta^0_{ijk}$ [degree] | $k^a_{ijk}$ [kJ mol$^{-1}$ rad$^{-2}$] |
|---|---|---|
| C-C-C | 112.7 | 488.273 |
| C-C-H | 110.7 | 313.8 |
| H-C-H | 107.8 | 276.144 |
| C-C-O | 109.5 | 418.4 |
| C-O-$H_O$ | 108.5 | 460.24 |
| H-C-O | 109.5 | 292.88 |
| C-C-C' | 111.1 | 527.184 |
| C'-C-H | 109.5 | 292.88 |
| C-C'-H' | 115.0 | 292.88 |
| C-C'-O' | 120.4 | 669.44 |
| H'-C'-O' | 123.0 | 292.88 |

**Table 5.** Dihedral angle torsion force constants. Dihedral angle torsion in the OPLS-AA force field is given as the first three terms of a Fourier series: $V(\varphi_{ijkl})=1/2(F_1(1+\cos\varphi_{ijkl})+F_2(1-\cos2\varphi_{ijkl})+F_3(1+\cos3\varphi_{ijkl}))$, where $\varphi_{ijkl}$ is the angle between $ijk$ and $jkl$ planes, and $\varphi_{ijkl}=0$ corresponds to the 'cis' conformation ($i$ and $l$ are on the same side).

| Dihedral type | $F_1$ [kJ mol$^{-1}$] | $F_2$ [kJ mol$^{-1}$] | $F_3$ [kJ mol$^{-1}$] |
|---|---|---|---|
| H-C-C-H [a] | 0 | 0 | 1.2552 |
| H-C-C-C [a] | 0 | 0 | 1.2552 |
| C-C-C-C [a] | 5.4392 | -0.2092 | 0.8368 |
| H-C-O-H$_O$ | 0 | 0 | 1.8828 |
| H-C-C-O | 0 | 0 | 1.9581 |
| C-C-O-H$_O$ | -1.4895 | -0.728016 | 2.058528 |
| C-C-C-O | 7.1588 | -2.092 | 2.773992 |
| H'-C'-C-C | 0 | 0 | 0 |
| H'-C'-C-H | 0 | 0 | 1.50624 |
| O'-C'-C-C | -1.159 | 5.137952 | -2.903696 |
| O'-C'-C-H | 0 | 0 | 0 |
| C'-C-C-C | -7.1002 | -1.907904 | 2.44764 |
| C'-C-C-H | 0 | 0 | -0.317984 |

[a] These parameters of the original OPLS-AA force field[5] were modified according to Reference.[7]

**Table 6.** Positions of the first intermolecular maxima for selected $g_{ij}(r)$ curves obtained by MD simulations (in Å).

| $i-j$ | $r$ [Å] |
|---|---|
| O-H$_O$ | 1.9 |
| O-H' | 2.8 |
| O'-H$_O$ | 1.9 |
| O'-H' | 2.75 |
| O-O | 2.8 |
| O-O' | 2.85 |
| O'-O' | 4.4 |